\pdfoutput=1

\documentclass[cits]{JINST}
\usepackage{amsmath, amsthm, amssymb}
\usepackage{mdwlist}
\usepackage{textcomp}
\usepackage{graphicx,subfigure}
\usepackage{fontenc}
\usepackage{pbox}
\usepackage{lineno}
\usepackage{url}
\usepackage[]{units}
\usepackage{xspace}

\newcommand{\teo}{TeO$_2$}
\newcommand{\znse}{ZnSe}
\newcommand{\znmo}{ZnMoO$_4$}
\newcommand{\tio}{TiO$_2$}
\newcommand{\halfo}{HfO$_2$}

\title{Characterization of single layer anti-reflective coatings for bolometer-based rare event searches}

\author{E.V.~Hansen$^{a}$, N.~DePorzio$^{b}$, \& L.~Winslow$^{c}$\\
\llap{$^a$}Drexel University, Department of Physics, Disque Hall Rm 816
32 South 32nd Street, Philadelphia, PA 19104\\
\llap{$^b$}Northeastern University, Department of Physics, 360 Huntington Ave., 111 Dana Research Center, Boston, MA 02115, USA\\
\llap{$^c$}Massachusetts Institute of Technology, Department of Physics, 77 Massachusetts Avenue, Cambridge, MA 02139, USA\\

E-mail: $^a$\email{evh32@drexel.edu}, $^c$\email{lwinslow@mit.edu}}
 
 \abstract{Combining analysis from phonon signals and photon signals is a powerful technique for reducing backgrounds in bolometer-based rare event searches. Anti-reflective coatings can significantly increase the performance of the secondary light-sensing bolometer in these experiments. As a first step toward these improvements, coatings of SiO$_2$, HfO$_2$, and TiO$_2$ on Ge and Si wafers were fabricated and characterized at room temperature and multiple angles of incidence. }
 

\newcommand{\vbb}{$0\nu\beta\beta$}
\begin{document}

\section{Introduction}\label{intro}
Rare-event searches are being pursued to answer some of the greatest mysteries in physics at the present time, namely: the nature of dark matter through direct detection (DM) and the possible Majorana nature of the neutrino through searches for neutrinoless double-beta decay (\vbb). In these experiments,  combining multiple signals is a powerful active background rejection technique. Scintillating bolometers use the combination of phonon and photon signals to discriminate between particle types. The CUPID (CUORE with Upgraded Particle IDentification)
~\cite{TheCUPIDInterestGroup2015CUPID:IDentification} 
and CRESST
~\cite{Angloher2016ResultsDetector, Kiefer2016In-situSearch} detectors are pursuing this technology for \vbb~and DM searches respectively.

\begin{figure}[t]
\begin{center}
\includegraphics[scale=0.3]{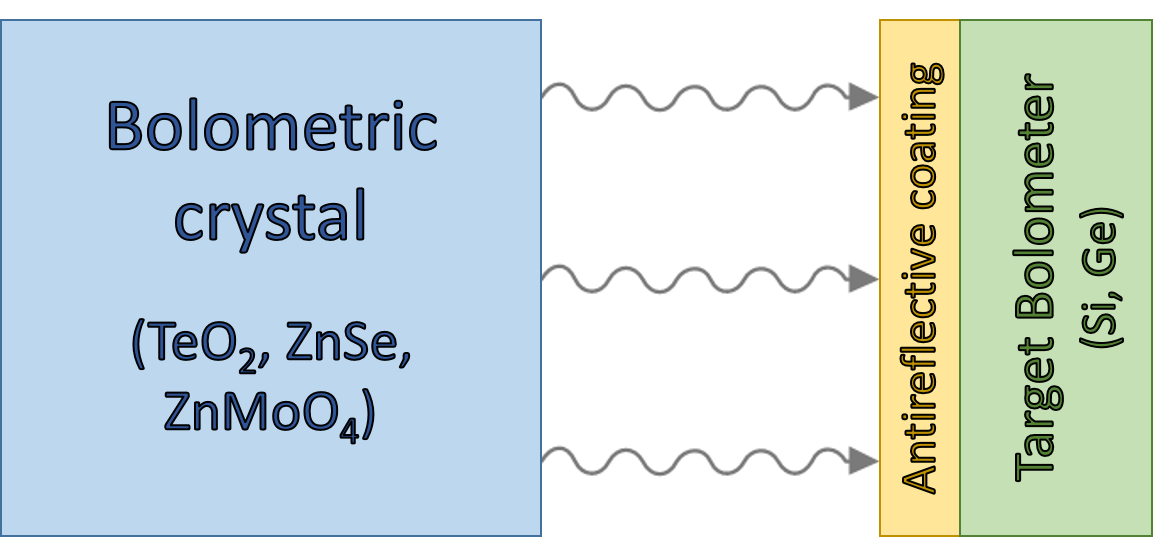}
\caption[]{An anti-reflective coating is deposited onto the auxiliary bolometer to improve transmission of Cherenkov or scintillation photons from the bolometric crystal. \label{cartoon}}
\end{center}
\end{figure}

A scintillating bolometer measures a phonon signal: the change in temperature in a crystal due to the interaction of charged particles with the crystal lattice. These interactions also produce a photon signal: scintillation light which is detected by a target Ge or Si bolometer. An anti-reflective coating on the target bolometer, as shown in Figure~\ref{cartoon}, increases light collection and therefore improves the energy resolution of the light measurement. In this paper, we discuss the optimization of an anti-reflective coating for two promising scintillating crystals containing \vbb~isotope: \znse~and \znmo. We also discuss optimizing the anti-reflective coating to detect Cherenkov light in non-scintillating crystals like \teo, the current CUORE crystal.

\section{Light Emission and Simulation}\label{sec::light}
\begin{figure}[t]
\begin{center}
\subfigure[bare Si]{\includegraphics[scale=0.23]{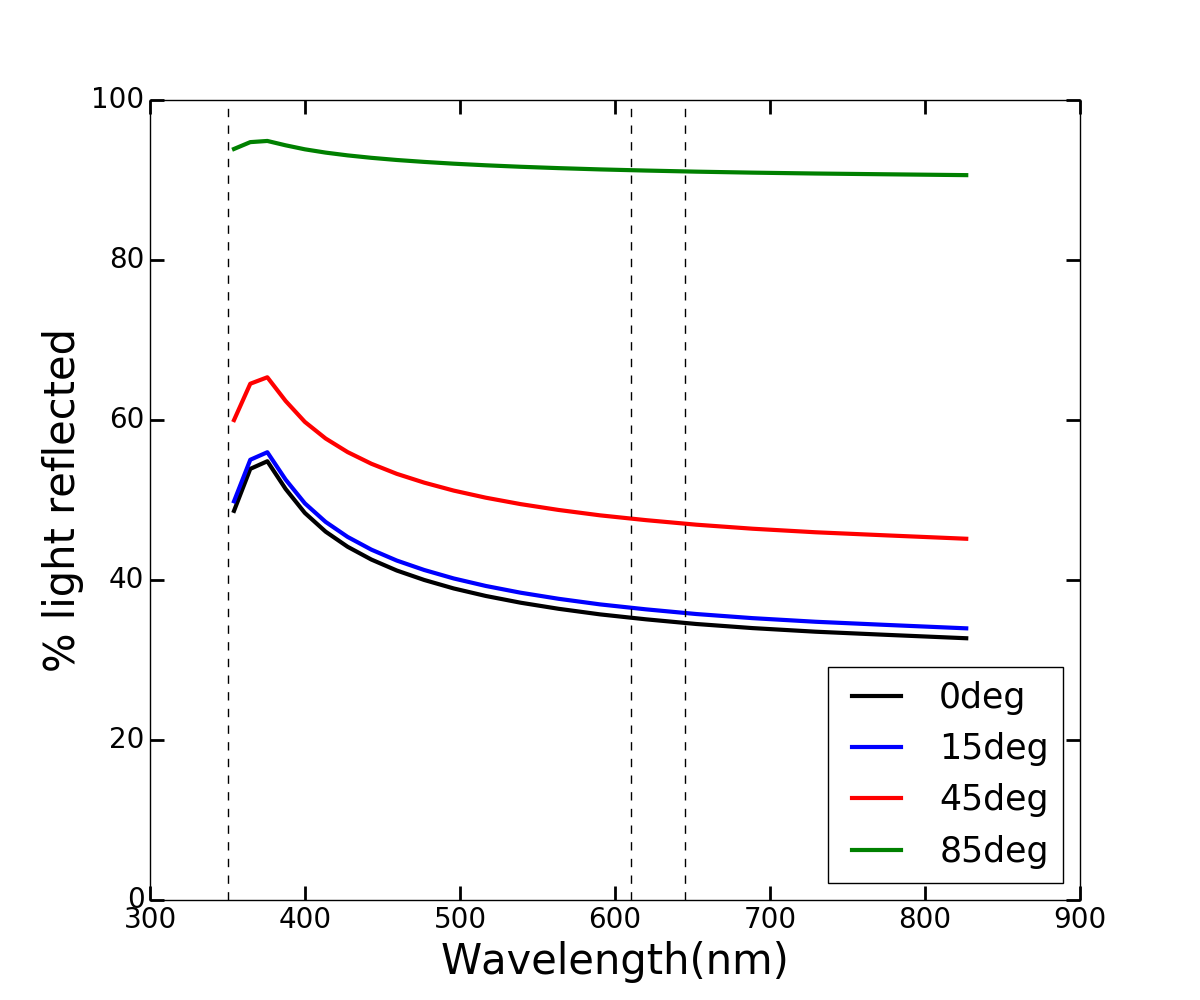}}
\subfigure[bare Ge]{\includegraphics[scale=0.23]{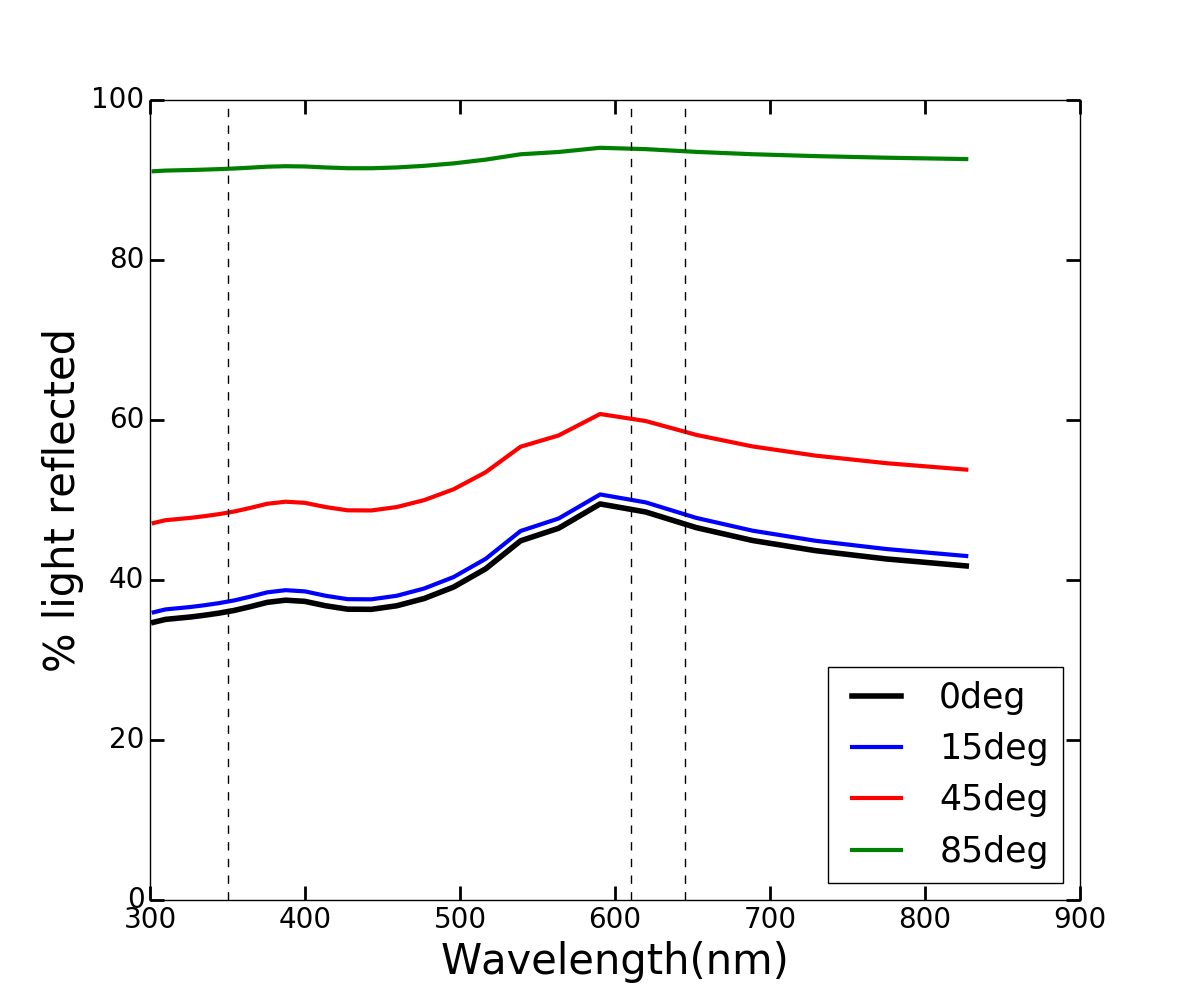}}
\caption[]{Reflectivity of light of various angles of incidence on bare substrates: (a) Si and (b) Ge~\cite{Aspnes1983DielectricEV}. Vertical dashed lines indicate the cutoff wavelength of TeO$_2$ (\unit[350]{nm}) and the peak wavelengths of scintillation for ZnMoO$_4$ (\unit[610]{nm}) and ZnSe (\unit[645]{nm}). For these wavelengths and normal incidence, bare Si reflects over 30\% of incoming light; bare Ge reflects over 40\%.\label{Ref_bare}}
\end{center}
\end{figure}

The scintillation spectra of \znse~and~\znmo~are well characterized at temperatures down to \unit[8]{K}. The spectra peak at \unit[645]{nm} and \unit[610]{nm} respectively~\cite{Mikhailik2010PerformanceTemperatures}. The absorption cutoff for \teo~is \unit[350]{nm}~\cite{Al-AniAGlasses}, forming the effective peak of the Cherenkov spectrum. Our default target bolometer is composed of hyper-pure germanium (HPGe) thin slabs run at a standard operating temperature of \unit[15-20]{mK}. We also study silicon (Si) since it has equivalent performance at these operating temperatures and is widely available. Using refractive index data from~\cite{Aspnes1983DielectricEV}, we find Ge substrates reflect $\sim$50\% of normal incident light while Si substrates reflect $\sim$35\%; see Figure~\ref{Ref_bare}.  The results for \znse~scintillation at \unit[645]{nm} and \teo~Cherenkov light are similar. 

\begin{figure}
\begin{center}
\includegraphics[width=0.5\textwidth]{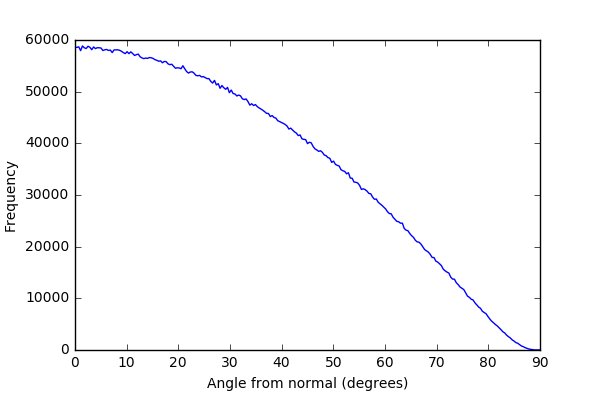}
\caption{Results of a GEANT4 Monte Carlo to measure the angle of incidence of photons generated by beta particles from a \teo\ bolometer. The angle of incidence is plotted with respect to normal incidence (0). This clearly demonstrates the failure of a normal incidence assumption. \label{angulardistribution}}
\end{center}
\end{figure}

As shown in Figure~\ref{Ref_bare}, the angle of incidence is critical for understanding the response of the target bolometer. A GEANT4\cite{geant1, geant2,geant3} Monte Carlo was constructed to examine the incidence angles of photons produced from beta particles distributed isotropically in position, angle, and energy throughout a \znmo\ bolometer. The light produced from these events was allowed to scatter within the crystal until it attenuated, escaped, or struck a target bolometer 1mm away from a single face. The relevant physical and optical properties of these materials, if not specifically included in the Geant4 NIST database, were obtained from literature\cite{Berg2014,Beeman2012,Green2008}. Results are shown in Figure~\ref{angulardistribution}. The average angle of incidence was 32$^\circ$ with a standard deviation of 21$^\circ$. This is a sizable deviation from normal incidence, and must be considered when selecting a coating/bolometer combination. In future work, the results of these measurements for the final target bolometers can be used either as input to or a benchmark of the Monte Carlo. 

%If the coating meets other design requirements for low background bolometer experiments, especially low radioactivity and robustness through thermal cycling (which will be examined in future work), then these numbers indicate that an anti-reflective coating could significantly improve the energy resolution of the target bolometer. 

\section{\label{CoatingInfo}Anti-Reflective Coatings}
As seen in Figure \ref{cartoon}, the target bolometer can be supplemented by depositing an anti-reflective coating to create a thin film structure where layers of contrasting refractive indices produce destructive interference in reflected beams and constructive interference in transmitted beams. This results in an overall increase of transmission. The performance of this structure depends heavily on the light's incident angle, wavelength, and polarization. The reflection coefficient for unpolarized light incident on a single layer anti-reflective coating (SLAR) is written~\cite{Tompkins1993AEllipsometry} as:
\begin{equation}\label{ref_first}R = \frac{1}{2} (|r_s|^2 + |r_p|^2)\end{equation}
\begin{equation}\label{ref_second}r_s=  \frac{r_{01,s} + r_{12,s}\exp(-2i\beta)}{1+ r_{01,s}r_{12,s}\exp(-2i\beta)} \hspace{0.5in} r_p =  \frac{r_{01,p} + r_{12,p}\exp(-2i\beta)}{1+ r_{01,p}r_{12,p}exp(-2i\beta)}\end{equation}
where the subscripts $s$ and $p$ refer to waves parallel or perpendicular to the plane of incidence and the numerical subscripts refer to the medium in which the light is traveling: 0 for the outside medium (ideally vacuum), 1 for the SLAR, and 2 for the substrate. If $\theta_{\text{inc}}$ is the angle of incidence onto the SLAR of thickness $d$, the phase shift is given by $\beta = 2\pi d N \cos\theta_{\text{inc}} / \lambda$. The individual reflection coefficients are given by
\begin{equation}r_{jm,s} =  \frac{N_m \cos\theta_j - N_j \cos\theta_m}{N_m \cos\theta_j + N_j \cos\theta_m} \hspace{0.5in} r_{jm,p} =  \frac{N_j \cos\theta_j - N_m \cos\theta_m}{N_m \cos\theta_j + N_j \cos\theta_m}\end{equation}
\vspace{0.2in}
and
\begin{equation}\label{ref_last} N_j = (n_j + ik_j) \hspace{0.5in} \theta_j = \arcsin\Big(\frac{N_0}{N_j}\sin\theta_{\text{inc}}\Big)\end{equation}
where $n$ and $k$ are the real and imaginary parts, respectively, of the complex refractive index of the material and are also dependent on wavelength. The ideal coating will minimize $R$ with respect to $n$ and $k$. %The ideal coating combination will also contribute minimally to system radioactivity and will perform well under thermal cycling.
%Subscript $Ge$ denotes the germanium wafer with $n_{Ge} = 5.48$ and $k_{Ge} = 0.823$.%
% Section 2 It has to be explicit the importance of Radiopurity in the choice of materials. This could be referred just before 2.2 Choice of substrates, by  clarifying that the condition of Radiopurity has to be applied to both, Substrates and Coatings.

\begin{table}[b]
\centering
\caption{Refractive index and total internal reflection angles for bolometer materials. At or above this angle from normal to the surface, a photon cannot exit the crystal into the vacuum.  The asterisk indicates that few data points exist for $\lambda$ in the ROI.}
\label{table_TIR}
\begin{tabular}{p{2cm} p{4cm} p{1cm}}
	Material & $n$ ($\lambda$) & $\theta_{\text{crit}}$ \\ \hline
	ZnSe & 2.58 (\unit[645]{nm})~\cite{Connolly1979SpecificationsMaterial} & 22.8$^{\circ}$\\
	ZnMoO$_4$ & $\sim$1.90 (\unit[655]{nm})$^\star$~\cite{Chernyak2013NuclearSearch}& 31.7$^{\circ}$\\
	TeO$_2$ & 2.25 (\unit[645]{nm})~\cite{Uchida1971OpticalTeO2}& 26.4$^{\circ}$\\
\end{tabular}
\end{table}
\subsection{Comparison to Previous Experiment (\emph{Mancuso, Beeman, et.al.})}
This work builds upon the study done by Mancuso et al.~\cite{Mancuso2014AnBolometersb}. They tested SiO$_2$ films, \unit[70]{nm} thick and deposited on a Ge substrate using a sputtering technique. The films were evaluated at \unit[$\sim$10]{mK}. SiO$_2$ decreased reflectivity by 18-20\%~\cite{Mancuso2014AnBolometersb}. In this paper, we manufacture similar coatings but characterize them at room temperature so that we can study the angular effects.

In Mancuso et.al., the scintillation light is assumed to reach the target bolometer at normal incidence. As we show in Section~\ref{sec::light} and as they indicate in their paper, this is a crude assumption. From first principles, the refractive index in the crystals of interest, \znse, \znmo,~and \teo, is high enough to allow total internal reflection at a relatively low angle; see Table~\ref{table_TIR}. This implies that even a slight deviation from normal (in the plane of incidence) as the ray leaves the primary bolometer can cause the refracted ray to deviate significantly from normal as it strikes the target bolometer.
The distance between primary and target bolometers is small enough ($\leq$\unit[1]{cm}) that the angle of incidence on the target bolometer will range from 0$^{\circ}$ to roughly 88$^{\circ}$ in the plane of incidence (assuming a CUORE-sized 5cm$\times$5cm crystal). From Equations~\ref{ref_first}-\ref{ref_last}, it is clear that a phase difference between reflected waves from the film layer and reflected waves from the substrate layer is introduced whenever the incident angle is non-normal, leading to a substantial change in reflectivity for non-normal incidence.

\subsection{\label{choice_substrate}Choice of Substrates} 
Ge is the logical choice for the target bolometer because it absorbs photons better than Si in the visible range; the absorption coefficient of Ge is roughly two orders of magnitude greater ~\cite{Dash1955Intrinsic300K}. However, Si is more readily available and the manufacturing processes are extremely well understood. These include techniques that could improve anti-reflective coatings, such as micromachining and deposition masking~\cite{Rahman2015Sub-50-nmCells}. Finally, Ge has a specific heat $\sim$4.5 times greater than that of Si~\cite{Coron2004HighlyMK}. Because signal amplitude is inversely proportional to the heat capacity of the device, Si detectors can be made into much larger systems without sacrificing signal quality~\cite{Biassoni2015LargeEffect}. Considering these factors, we chose to study both Ge and Si as target bolometer candidates for this work. 

\subsection{\label{choice_coatings}Choice of Coatings}
For a particular substrate, the minimum possible reflectivity using a SLAR can be calculated using the curves from Equation~\ref{ref_first}. For a Ge substrate and incident \unit[645]{nm} light from the \znse\ scintillation peak, the complex refractive index of Ge has components $n=5.36$ and $k=0.705$. Assuming no losses to the thin film and normal incidence, the reflection coefficient can be decreased maximally with a coating medium with refractive index $n=2.32$. The closest match considering available materials is \tio; see Table~\ref{table_n_coatings}. \halfo~is also a possibility. Performing the same calculation for a Si substrate, the minimum reflectivity is achieved for a coating with $n=1.96$. Si$_3$N$_4$, HfO$_2$, and Al$_2$O$_3$ (sapphire) are good matches. 

\begin{table}
\centering
\caption{refractive index at 645nm for coatings considered in this work.}
\label{table_n_coatings}
\begin{tabular}{p{2cm} p{2cm} p{4cm}}
Material & n (645nm) & Source\\ \hline
Al$_2$O$_3$ & 1.76	& Malitson (1972)~\cite{Malitson1972RefractiveSapphire} \\
HfO$_2$ 	& 2.10 	& Wood (1990)~\cite{Wood1990OpticalYttria}\\
Si$_3$N$_4$ & 2.00	& Philipp (1973)~\cite{Philipp1973OpticalNitride}\\
SiO$_2$ 	& 1.48 	& Gao (2013)~\cite{Gao2013RefractiveDesigns}\\
TiO$_2$ 	& 2.58 	& Devore (1951)~\cite{Devore1951RefractiveSphalerite}\\
\end{tabular}
\end{table}

A second anti-reflective layer can be used to further reduce reflectivity. The optimization proceeds similarly to above, except that, in this case, the refractive index is tuned to the first coating instead of the substrate. For example, a Ge-TiO$_2$ target system could be improved by adding second coating with a refractive index of $n=1.52$. GeO$_2$ ($n=1.60$ ~\cite{Fleming1984DispersionGlasses}) is not readily available, but the more common SiO$_2$ is a good candidate. Traditionally, SiO$_2$ has been used as a  single layer coating for Si substrates because it is readily available, easily manufactured, and has better mechanical strength and adhesion under thermal cycling. In this work, we focus on single layer coatings, but future work will include two layer systems.

\begin{figure}[!t]
\begin{center}
\subfigure[\ ]{\includegraphics[scale=0.23]{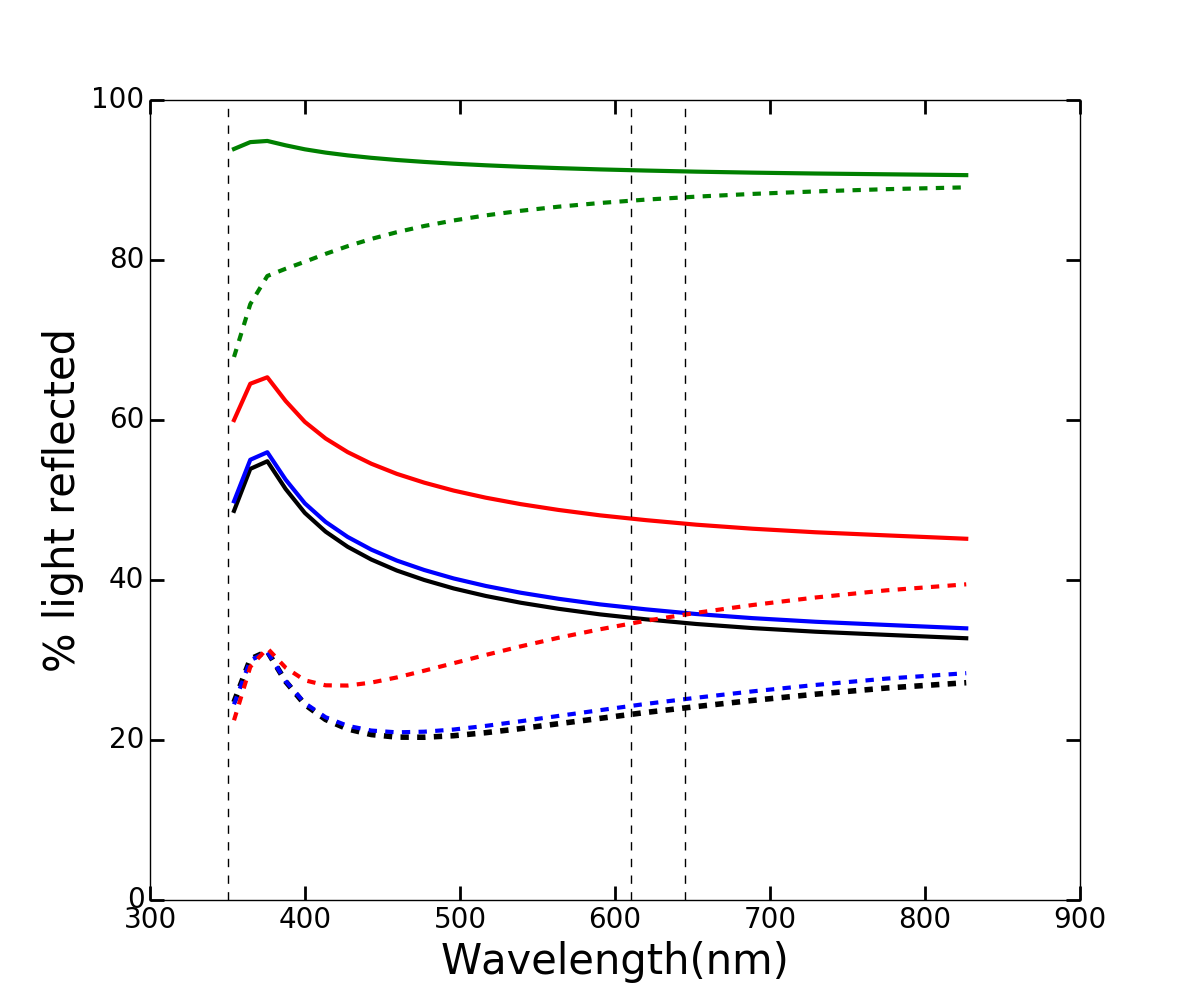}\includegraphics[scale=0.23]{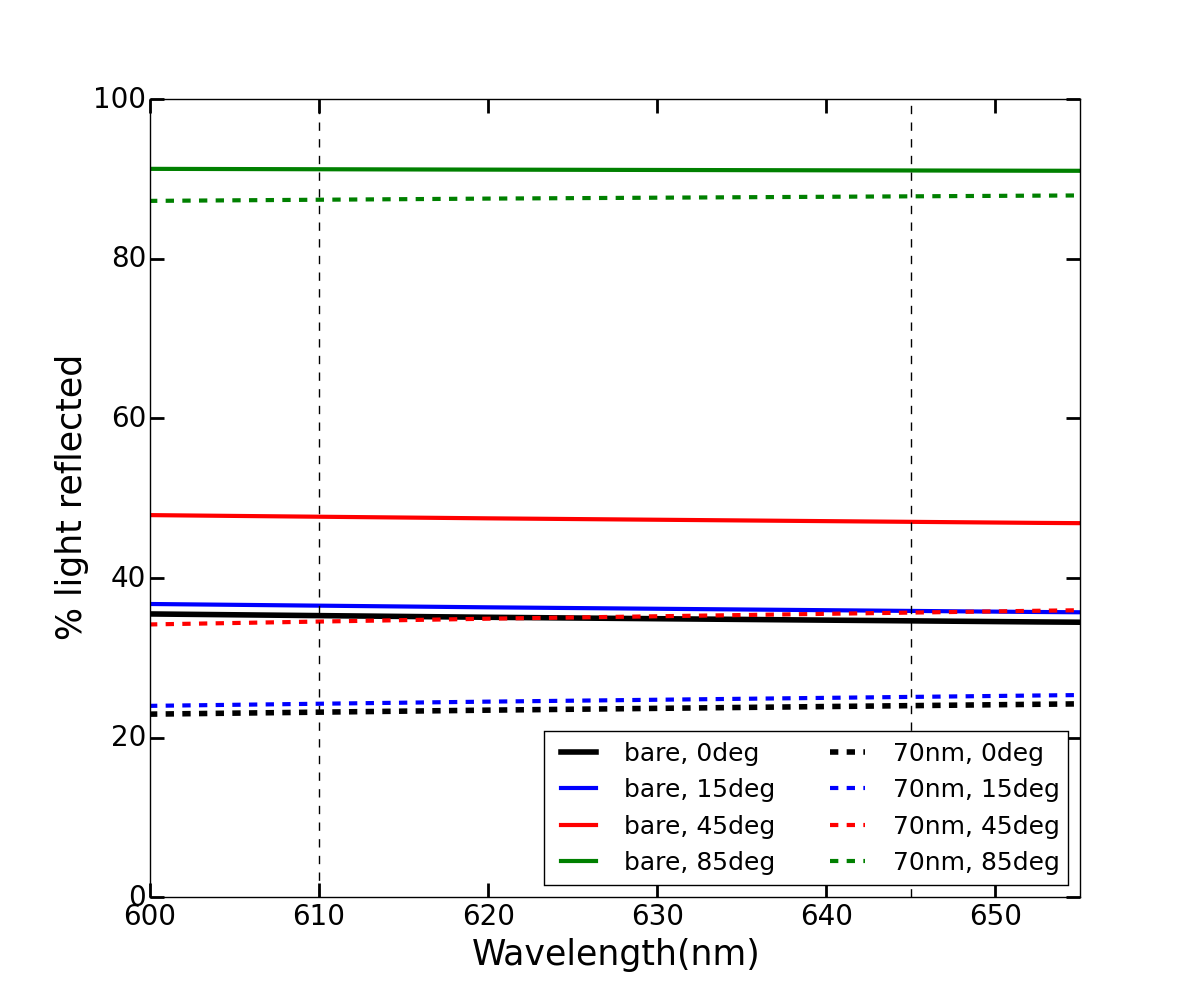}}
\subfigure[\ ]{\includegraphics[scale=0.23]{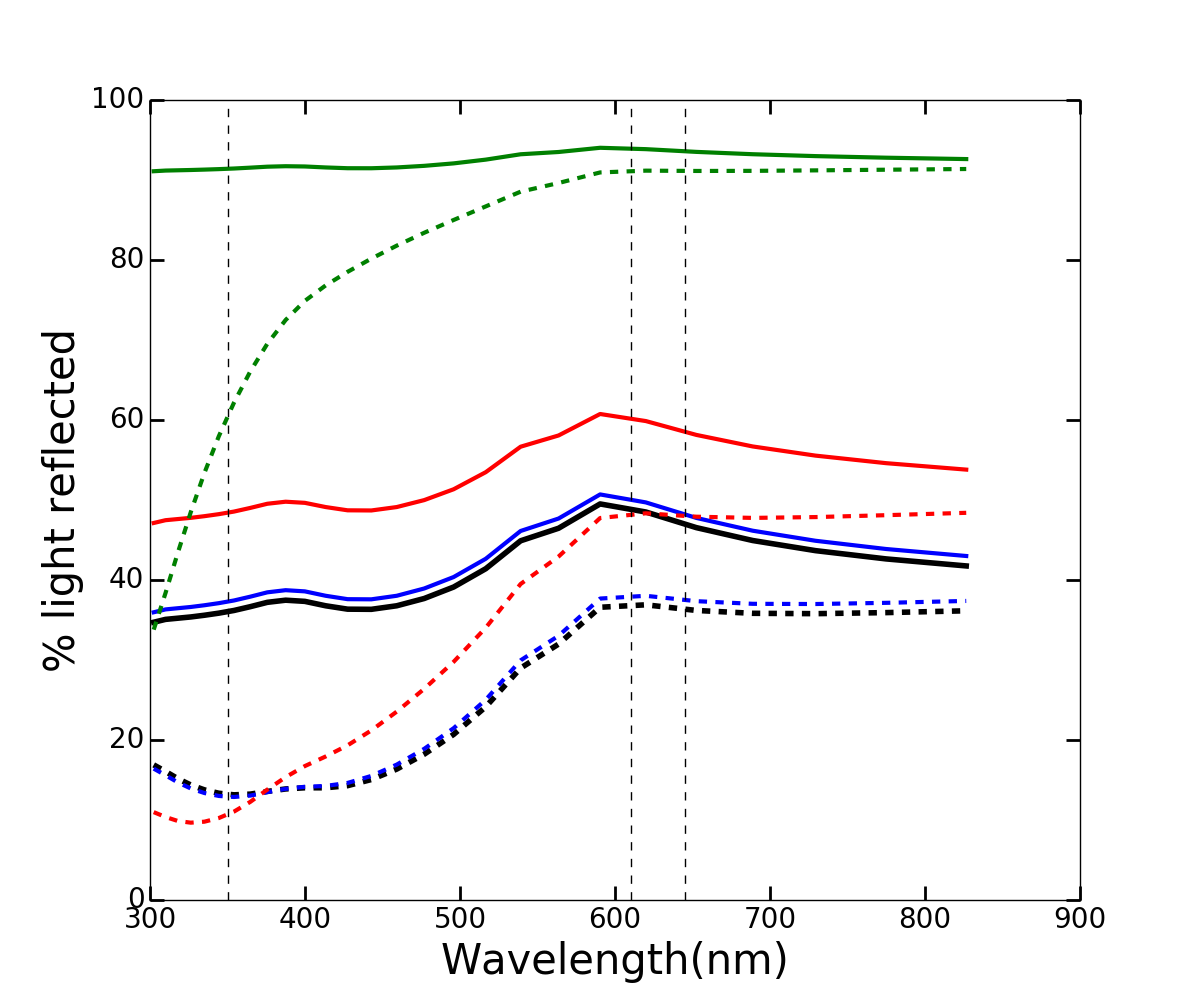}\includegraphics[scale=0.23]{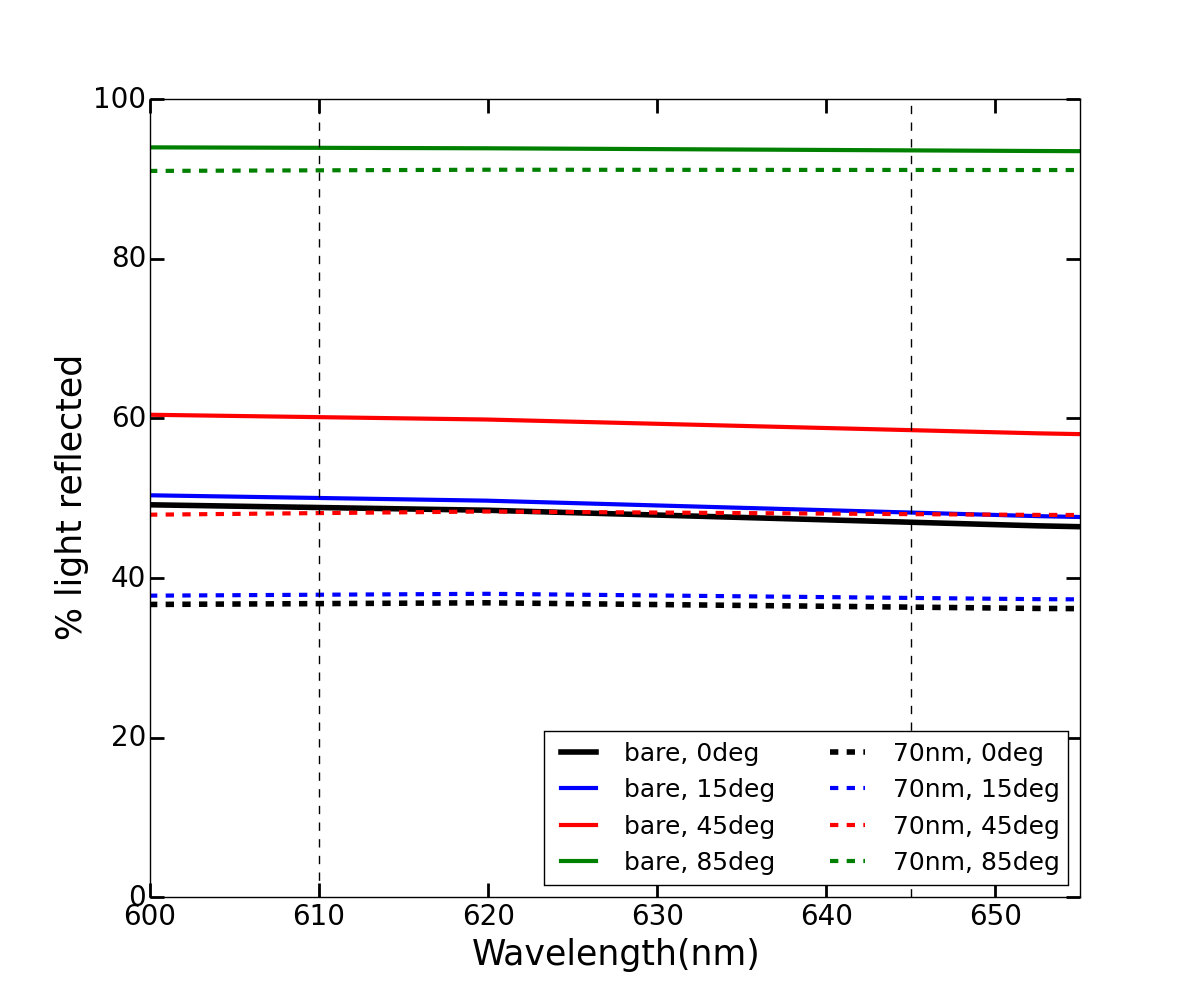}}
\caption[]{Dependence of reflection on angle of incidence for Si (a) and Ge (b). Values for bare substrate (solid) and \unit[70]{nm} SiO$_2$~\cite{Gao2013RefractiveDesigns} single layer antireflective coating (dashed) are presented. Plots on left emphasize the wavelengths of interest (\unit[350]{nm} cutoff wavelength for TeO$_2$ and \unit[610]{nm} and \unit[645]{nm} scintillation peaks of ZnMoO$_4$ and ZnSe, respectively).  Calculations are taken directly from equations \ref{ref_first}-\ref{ref_last} and contain no experimental data. \label{PredRefCurvesByAngle}}
\end{center}
\end{figure}

As previously noted, the performance of the anti-reflective coating is dependent on the angle of the incoming light. Using equation \ref{ref_first} for non-normal incidence, reflection curves were plotted for use in selecting SLAR candidates; these curves for SiO$_2$ on Si and Ge substrates can be seen in Figure~\ref{PredRefCurvesByAngle}. The coating always improves the performance of the target bolometer. However, the improvement may not be sufficient to overcome the increased complexity and cost of adding the coating.

The above calculations assume all reflections are specular, the refractive index is constant throughout the thickness of the SLAR, and there are no losses in the SLAR ($k_{\text{SLAR}}=0$). The incident light is also assumed to be unpolarized. This final assumption is not valid for \teo~due to birefringence. In this case, the refractive index varies between 1 and the true refractive index depending on the axis of propagation.

\begin{figure}
\begin{center}
\subfigure[Coated Si]{\includegraphics[scale=0.41]{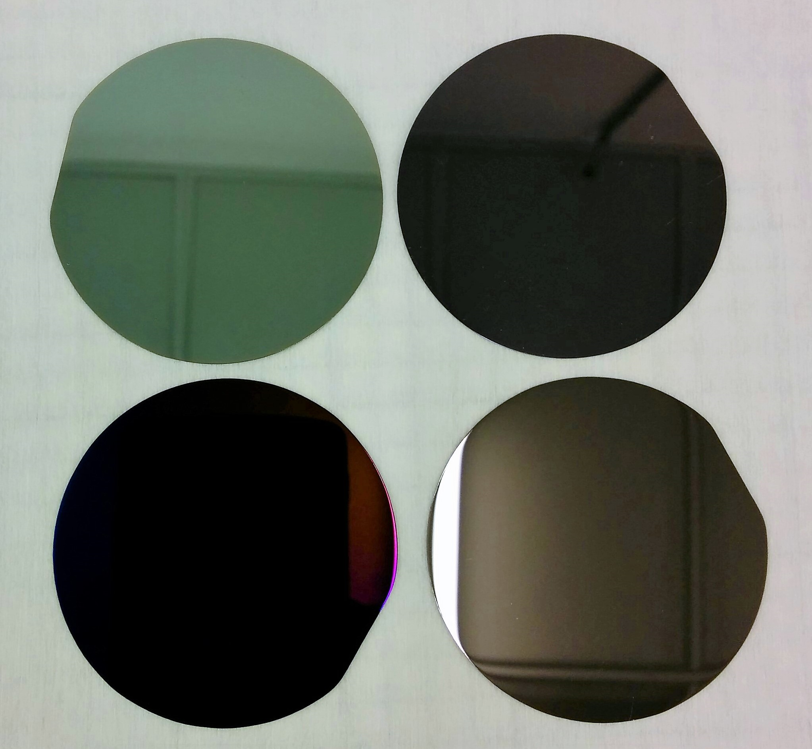}} \hspace{.3in}
\subfigure[Coated Ge]{\includegraphics[scale=0.41]{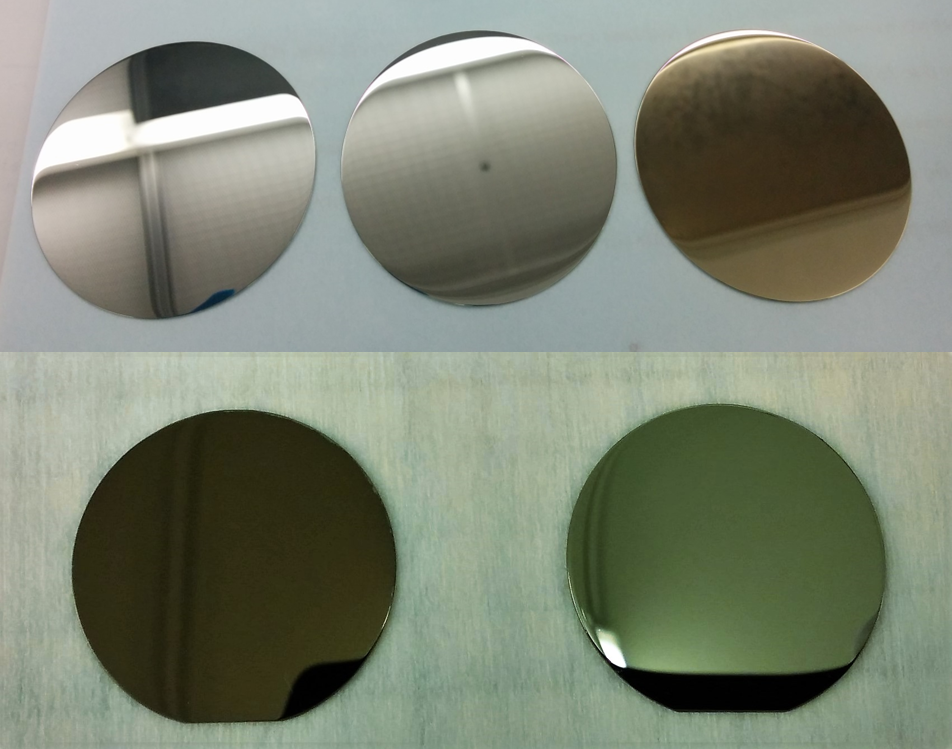}}
\caption[]{Wafers post-fabrication: (a) Si wafers clockwise from top left: SiO$_2$ \unit[180]{nm}, SiO$_2$ \unit[90]{nm},  SiO$_2$ \unit[60]{nm}, HfO$_2$ \unit[60]{nm}. (b) Ge wafers clockwise from top left: blank, Al$_2$O$_3$ \unit[7]{nm}, TiO$_2$ \unit[20]{nm}, SiO$_2$ \unit[70]{nm}, TiO$_2$ \unit[30]{nm}.\label{wafer_closeup}}
\end{center}
\end{figure}

Based on their refractive index and availability, the following coatings were selected. For Ge, we study Al$_2$O$_3$, SiO$_2$, TiO$_2$, and HfO$_2$. For Si, we study SiO$_2$ and HfO$_2$. Figure~\ref{wafer_closeup} shows the wafers produced for this study. Al$_2$O$_3$ is a promising candidate. We were not able to obtain the materials to deposit it on Si, so it was only used for Ge.

\begin{table}
\centering
\caption{Characteristics of wafers used in this work}
\label{table_wafer_info}
\begin{tabular}{p{2.5cm} p{3.5cm} p{3.5cm} p{3.5cm}}
			& Ge (1)				& Ge (2) 					& Si \\ \hline
Orientation	& crystalline, <100> 	& crystalline <100>			& crystalline, <100> \\
Thickness 	& \unit[500]{$\mu$m}	& \unit[350]{$\mu$m}		& \unit[280]{$\mu$m} \\
Diameter	& \unit[2]{in}			& \unit[1]{in}				& \unit[2]{in}\\
Doping		& undoped				& undoped 					& N Type, P Doped \\
Polish		& 2SP					& 1SP						& 1SP\\
Resistivity & \unit[>50]{$\Omega\cdot$cm} & \unit[30]{$\Omega\cdot$cm}	& \unit[1-100]{$\Omega\cdot$cm} (test)
\end{tabular}
\end{table}

\section{Other Requirements}
In addition to the target bolometer's optical properties, DM and \vbb~ requirements have stringent requirements on surface and bulk radioactivity contamination for detector components. A complete background model like that constructed for CUORE\cite{Alduino:2017qet} will set these specifications for the coating and substrate. The CUPID program already has extensive experience with the procurement and handling of Ge and Si wafers. When the final candidate coating are chosen, the bulk material will need to be counted to ensure that it meets the desired specifications. Devoted tests will also be needed for understanding the coatings robustness under thermal cycling. This is all part of future work.

\section{\label{nanolab}Fabrication}
The 1-inch Ge wafers and 2-inch Si wafers were procured from University Wafer~\cite{Http://www.universitywafer.com/}. The \unit[2]{in} Ge wafers were purchased from MTI Corporation~\cite{Http://www.mtixtl.com/}. Table~\ref{table_wafer_info} summarizes the properties of these wafers. The wafers were coated at the UCLA Nanoelectronics Research Facility (NRF) in a class-1000 multiuse cleanroom. %The wafers were packaged for use in clean room and kept in airtight sheathes so to not expose them to atmosphere outside of the clean room. 
% This isn't true. They were exposed to oxygen a couple of times because I had to take them to UCSB for testing and I'm pretty sure the box wasn't airtight.

\paragraph{Al$_2$O$_3$, HfO$_2$, TiO$_2$ coatings}
Non-silicate coatings were deposited using a Fiji thermal atomic layer deposition (ALD) system. In such a system, precursors were pulsed into an Argon atmosphere such that for each pulse a single atomic layer adhered to the surface of the wafer. Each wafer was loaded into the machine at room temperature and atmosphere, and processed at \unit[200]{$^{\circ}$C} and \unit[0.02]{mTorr}. 

Al$_2$O$_3$ was used as a test wafer. It is a common coating that starts from TMA (trimethylaluminum) or Al(CH$_3$)$_3$. For HfO$_2$ coatings, precursors of Hf(NMe$_2$)$_4$ and H$_2$O were pulsed at \unit[0.06]{sec} each until the desired thickness was reached. For TiO$_2$, coatings were processed similarly from a precursor of Tetrakis(Dimethylamido)Titanium (Ti(NMe$_2$)$_4$). 

The ALD process was lengthy and required up to six hours to produce a single wafer. This technique has the advantage that more complex coating geometries can be achieved through nanopatterning. These geometries will be explored in future work.

\paragraph{SiO$_2$ coatings}
Silicate coatings were deposited using a High Deposition BMR plasma-enhanced chemical vapor deposition (PECVD). The system uses time-varying magnetic fields to generate highly dissociated plasmas of the precursor material which allows for a higher rate of deposition. Precursors of SiH$_4$ and O$_2$ gasses were used to create SiO$_2$ films at rates of up to \unit[3000]{{\AA}/min}. In the case of the BMR PECVD, the input is a desired time of deposition (as opposed to a desired thickness). This led to less precision in the final thickness of the coating; however, this process was very efficient and several wafers of different thicknesses could be produced over the course of a few hours. 

%E - Need better labels here, per Kelley

\section{\label{ellipsometry}Characterization}
\begin{figure}[!tbp]
\begin{center}
\includegraphics[width=0.5\textwidth]{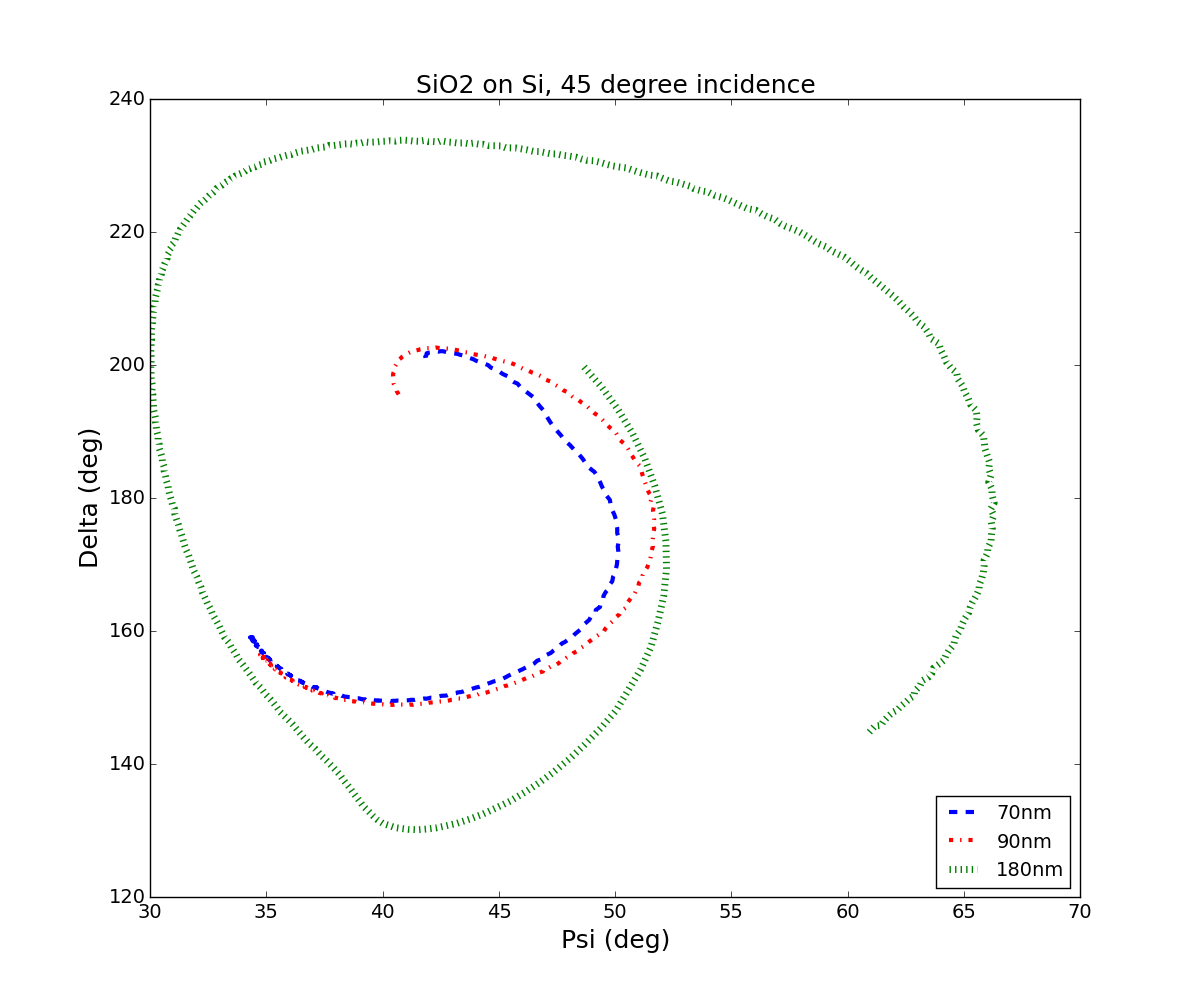}
\caption[]{Example of `raw' ellipsometric data $\Psi$ and $\Delta$ (defined in Section \ref{ellipsometry}). These values are then modeled using Cauchy film theory to find indexes of refraction shown in the next figure. \label{psideltaexp}}
\end{center}
\end{figure}

Samples were characterized using fixed and variable angle ellipsometry. The coated substrates were subjected to unpolarized light at an angle which reflected into a detector to measure the relative amplitude and phase of s- and p-polarizations. Relative amplitudes and phases of polarizations are described as a function of wavelength and angle by two variables ~\cite{Tompkins1993AEllipsometry}:
\begin{equation}\label{psidelta_first}
\tan\psi = \frac{|r_p|}{|r_s|} \hspace{0.2in} \text{and} \hspace{0.2in} \Delta = \delta_{rp} - \delta_{rs} 
\end{equation}
which can be combined to measure the total ratio of polarized reflections.
\begin{equation}\label{psidelta_second}
\tan\psi\exp(-i\Delta) = \frac{r_p\ (N_1,N_2,\theta_{inc}, d)}{r_s\ (N_1,N_2,\theta_{inc},d)}
\end{equation}
Figure~\ref{psideltaexp} shows an example of representative $\Psi/\Delta$ data. 

\begin{table}[!tbp]
\centering
\caption{Samples characterized for this study. Other samples were manufactured but their models did not reach acceptable goodness of fit values and therefore are not presented here. Discrepancies between ULVAC and Woolam measurements can be attributed to separate calibrations.}
\label{table_coatinginfo}
\begin{tabular}{lllll}
\\
Sample \# & Substrate & Coating & \parbox{3cm}{Thickness (nm)\\(ULVAC)} & \parbox{3cm}{Thickness (nm) \\ (Woolam)} \\ 
\hline
1 & Si		& SiO$_2$		& 66.116 $\pm$ 1.899	&  75.655 $\pm$ 0.021    \\
2 & Si      & SiO$_2$       & 79.627 $\pm$ 1.245    &  90.09 $\pm$ 0.49   \\
3 & Si      & SiO$_2$       & 181.902 $\pm$ 0.782	&  180.029 $\pm$ 0.037 \\
4 & Ge		& SiO$_2$		& --					&  75.26 $\pm$ 2.294 \\
5 & Ge		& SiO$_2$		& --					&  140.84 $\pm$ 3.13
\end{tabular}
\end{table}

At this point, refractive index and thickness information for the coating was modeled and fit to experimental curves using equations \ref{ref_second}-\ref{ref_last}. 
It was not possible to directly calculate total reflection from these measurements, but a satisfactory result was found by 
reusing the calculated refractive index and thickness in equations \ref{ref_first}-\ref{ref_last}.
It should be noted that the measurement error of $\psi$ and $\Delta$ are systematic; calculations of film thickness and optical characteristics are model dependent.

The UCLA Nanoelectronics Research Facility ULVAC UNECS-2000 fixed angle ellipsometer was used for immediate characterization of the Si wafers. Ge is less common, so no model was available. The UCSB Nanofabrication Facility's Woollam M2000DI VASE Spectroscopic Ellipsometer was used to fully characterize all samples  at variable angles.

\begin{figure}
\begin{center}
\subfigure[SiO$_2$ on Si]{\includegraphics[scale=0.23]{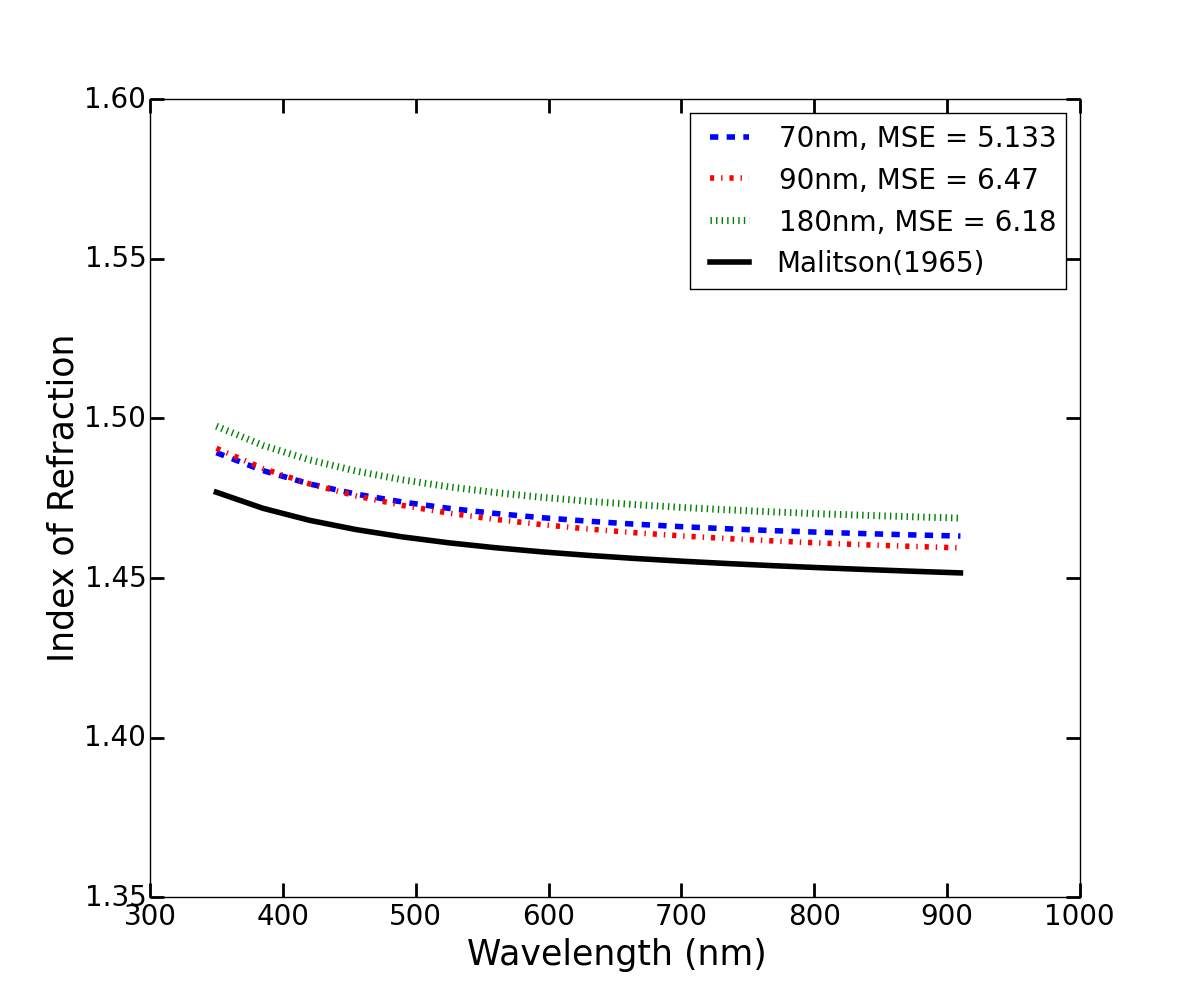}}
\subfigure[SiO$_2$ on Ge]{\includegraphics[scale=0.23]{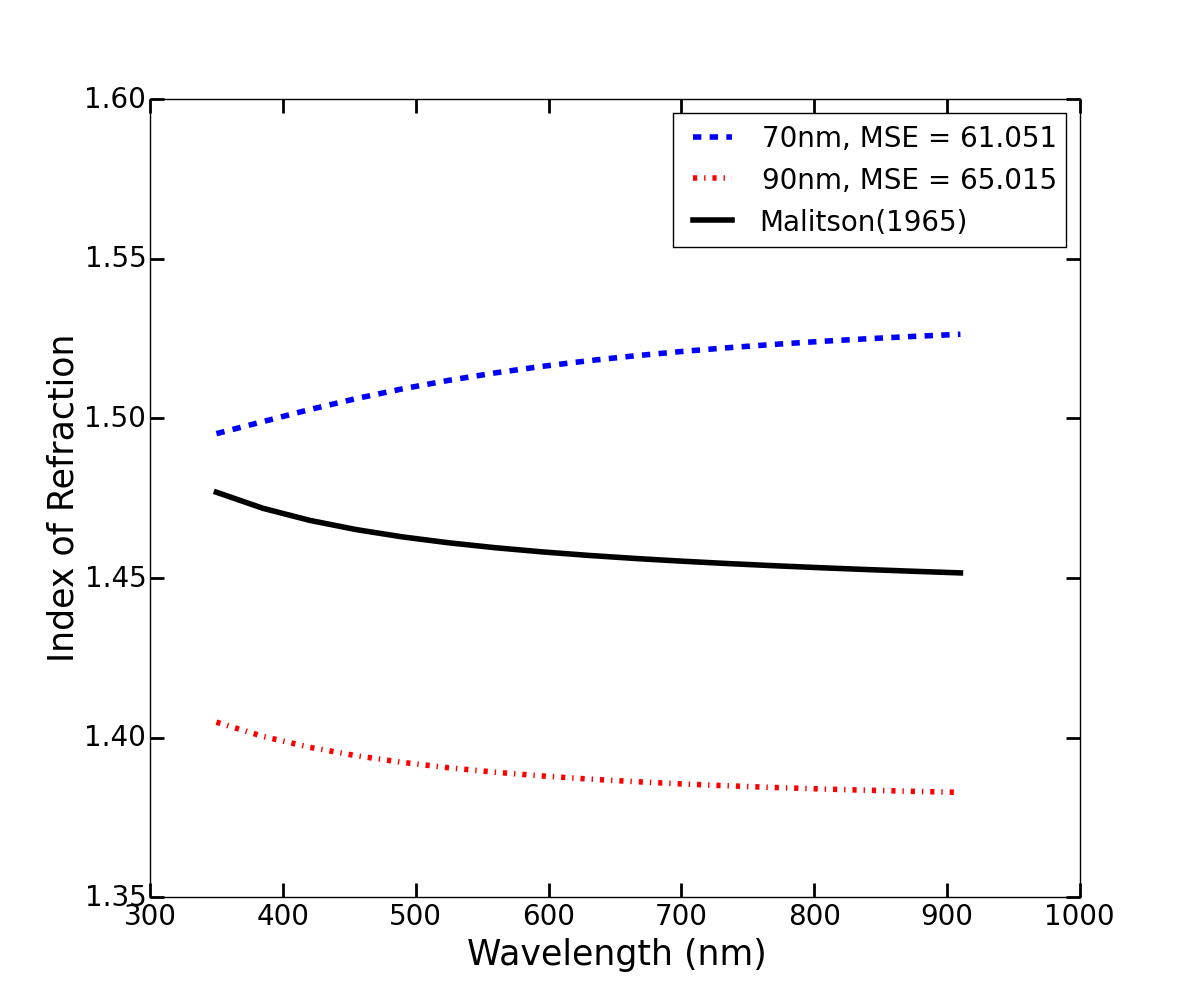}}
\caption[]{Refractive index fit for SiO$_2$ on (a) Si and (b) Ge. MSE values correspond to goodness of fit (see Eq~\ref{MSE}; MSE$<10$ are excellent, MSE$<100$ are acceptable). Black solid line indicates literature values. \label{NCurves}}
\end{center}
\end{figure}

%original:  Film thickness and single-point indices were determined using the ULVAC companion software for normal incidence, and the Woollam ellipsometer's companion software \emph{CompleteEase} was used for variable angle incidence. 
% rewrite:
Film thickness and single-point were analyzed in two groups. The ULVAC companion software was used for normal incidence, and the Woollam ellipsometer's companion software \emph{CompleteEase} was used for variable angle incidence. 
The results are shown in Table~\ref{table_coatinginfo}. SiO$_2$ coatings were assumed to follow Cauchy's equation:
\begin{equation}\label{cauchy}
n(\lambda) = A + \frac{B}{\lambda^2} + \frac{C}{\lambda^4} + ...
\end{equation}
The resultant fits for SiO$_2$ on Si and Ge are shown in Figure~\ref{NCurves}. The goodness of fit of these curves is demonstrated by the root mean squared error, defined as:
\begin{equation}\label{MSE}
\text{MSE} = \Big[\frac{1}{3n-m} \sum_{i=1}^n  \Big\{(N_{Ei} - N_{Gi})^2 + (C_{Ei} - C_{Gi})^2 + (S_{Ei} - S_{Gi})^2  \Big\}  \Big]^{1/2} \cdot 1000
\end{equation}
where subscripts $E$ and $G$ refer to measured and modeled parameters, respectively, and 
%E - list of fit parameters unclear? per Kelley
\begin{eqnarray*}
n &=& \text{number of wavelengths}\\
m &=& \text{number of fit parameters} = 3 \text{ (as follows)}\\
N &=& \cos(2\Psi)\\
C &=& \sin(2\Psi)\cos(\Delta)\\
S &=& \sin(2\Psi)\sin(\Delta)\\
\end{eqnarray*}
The error in the parameters $N$, $C$ and $S$ that is introduced by the measurements is estimated to be typically 0.001, leading to the extra multiplicative factor of "1000". From equation~\ref{MSE}, this implied that a fit with an MSE on the order of unity was considered in "perfect agreement" with the data. Values less than 100 were accepted for our purposes (as suggested by~\cite{A.WoollamCo.Inc2008CompleteEASEManual}). Fits for HfO$_2$ and TiO$_2$ samples had MSE values greater than 100, and were rejected. This method of testing goodness of fit is done automatically through the \emph{CompleteEase} software. We found that non-standard coatings require specific calibrations of the ellipsometer. In the future, this will be done in order to characterize the total reflectivity and transmissivity of additional SLAR

\begin{figure}[!h]
\begin{center}
\subfigure[$\theta_\text{inc}=0^\circ$]{\includegraphics[scale=0.23]{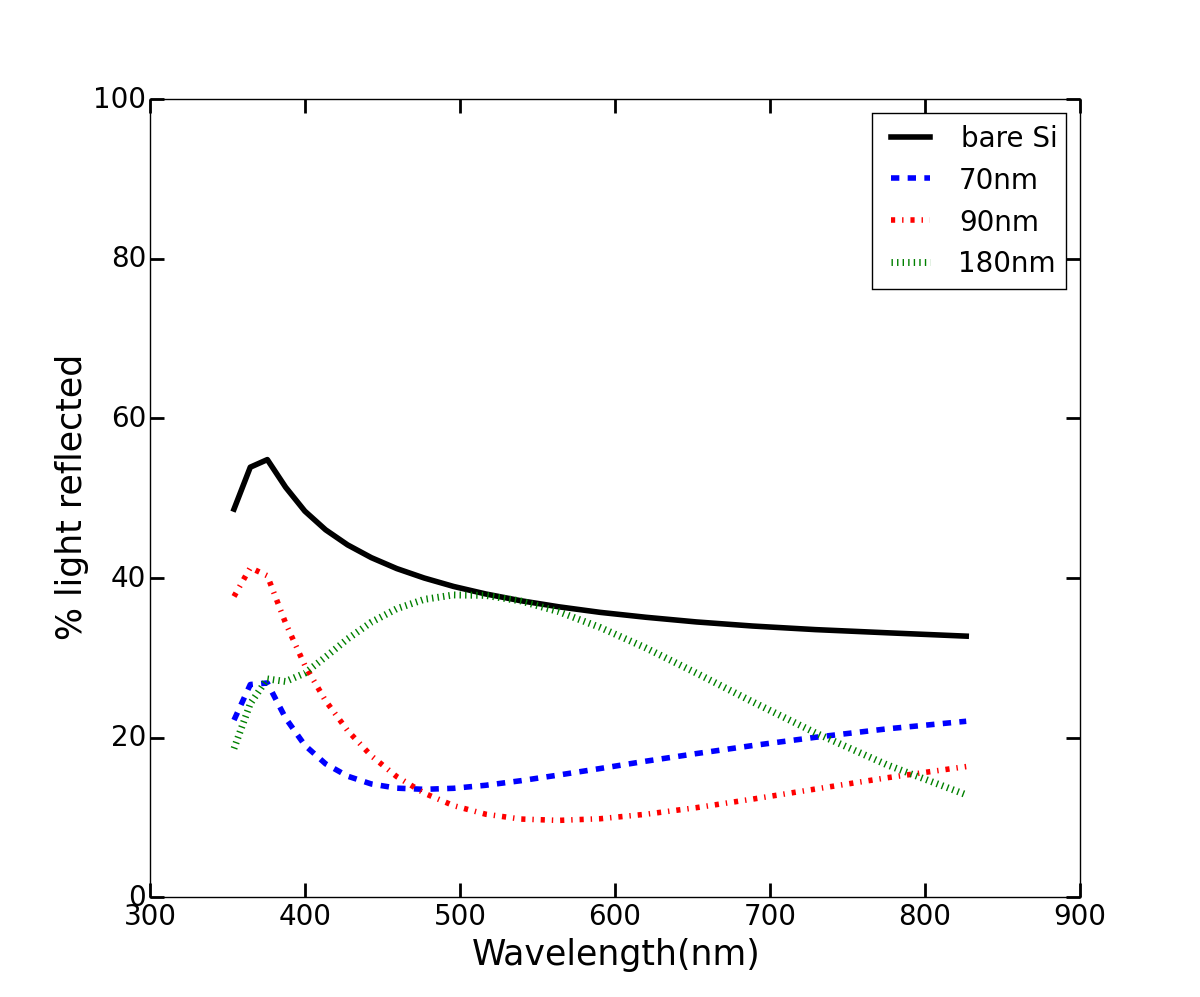}}
\subfigure[$\theta_\text{inc}=15^\circ$]{\includegraphics[scale=0.23]{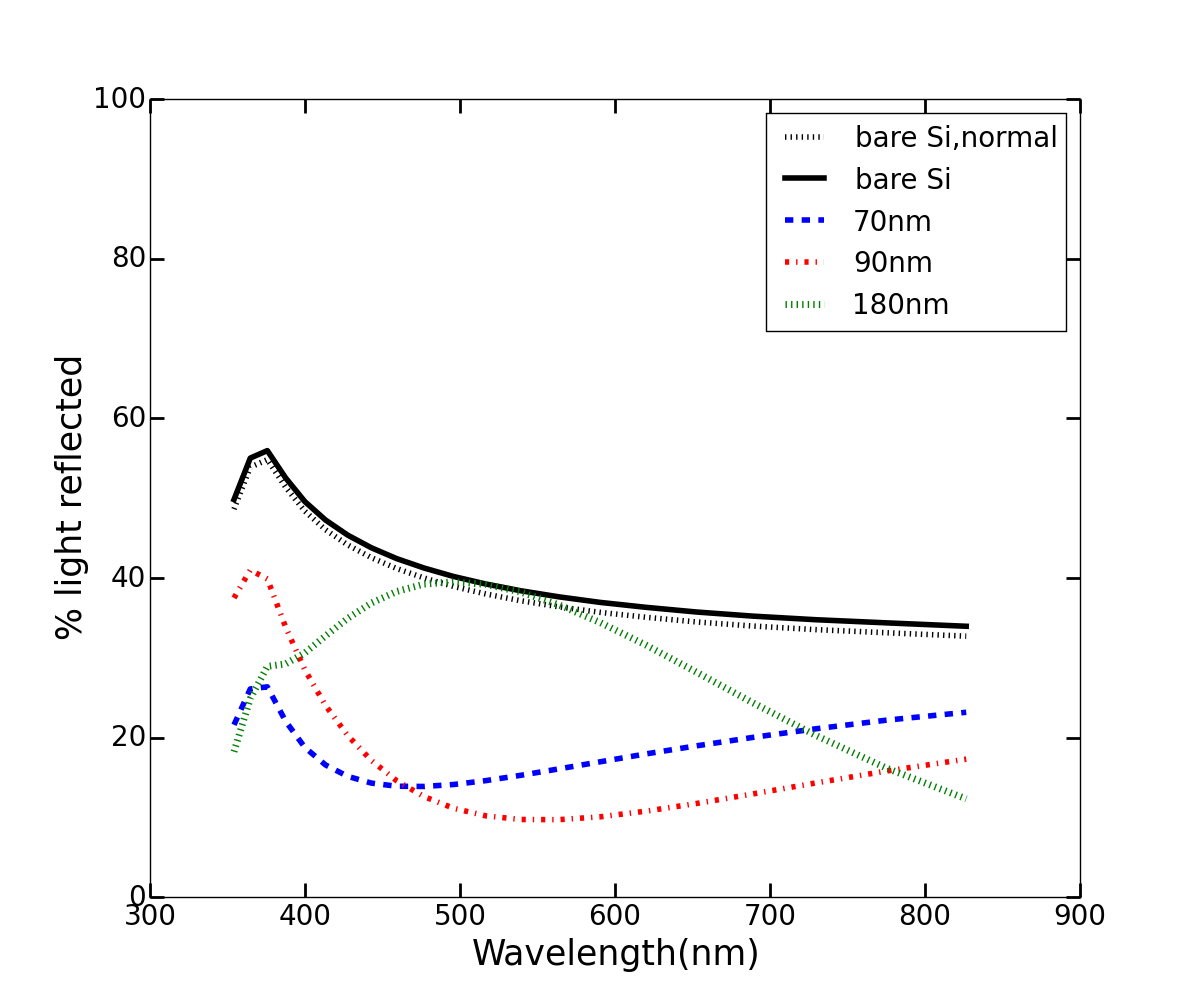}}
\subfigure[$\theta_\text{inc}=45^\circ$]{\includegraphics[scale=0.23]{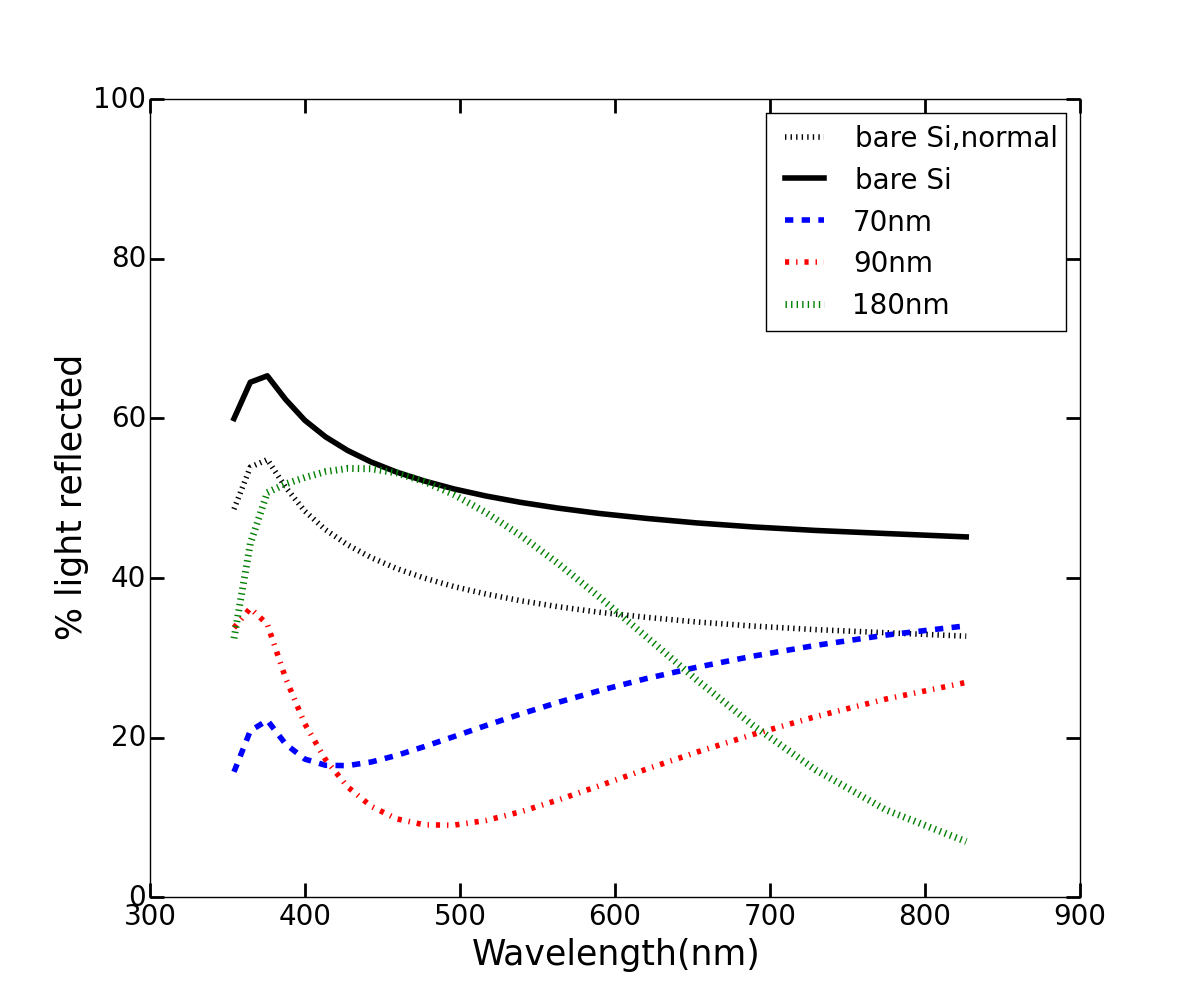}}
\subfigure[$\theta_\text{inc}=85^\circ$]{\includegraphics[scale=0.23]{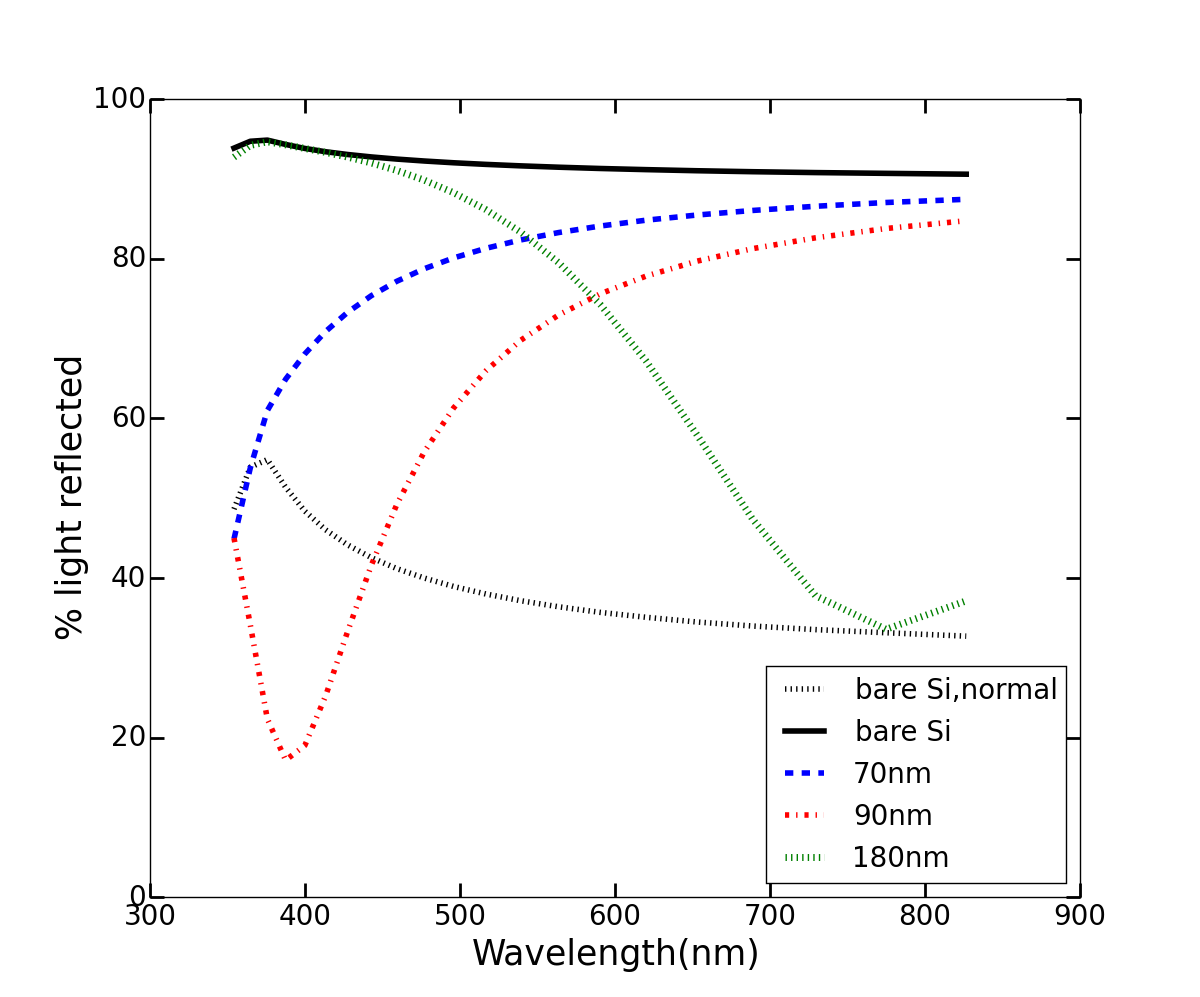}}
\caption[]{Predicted reflectivity values for SiO$_2$ coating on Si substrate for varied angles of incidence. Black solid curve indicates reflectivity of bare Si~\cite{Aspnes1983DielectricEV}. Angles of incidence are as follows: (a) 0$^\circ$, (b) 15$^\circ$, (c) 45$^\circ$, (d) 85$^\circ$.  Colored reflectivity curves are calculated based on ellipsometric measurements for $n(\lambda)$ (see Figure \ref{NCurves}). Thicknesses of 70, 90, and \unit[180]{nm} (with MSE values 5.13, 6.47 and 6.18 respectively) are displayed. For plots (b)-(d), the reflectivity curve for normal incidence is provided as a reference (black dotted). For all angles of incidence and thicknesses, coating with SiO$_2$ should improve reflectivity. Recommendations for wavelengths of interest can be found in Section \ref{PredRef}.\label{RefCurvesSiO2onSi}}
\end{center}
\end{figure}

\begin{figure}[!h]
\begin{center}
\subfigure[$\theta_\text{inc}=0^\circ$]{\includegraphics[scale=0.23]{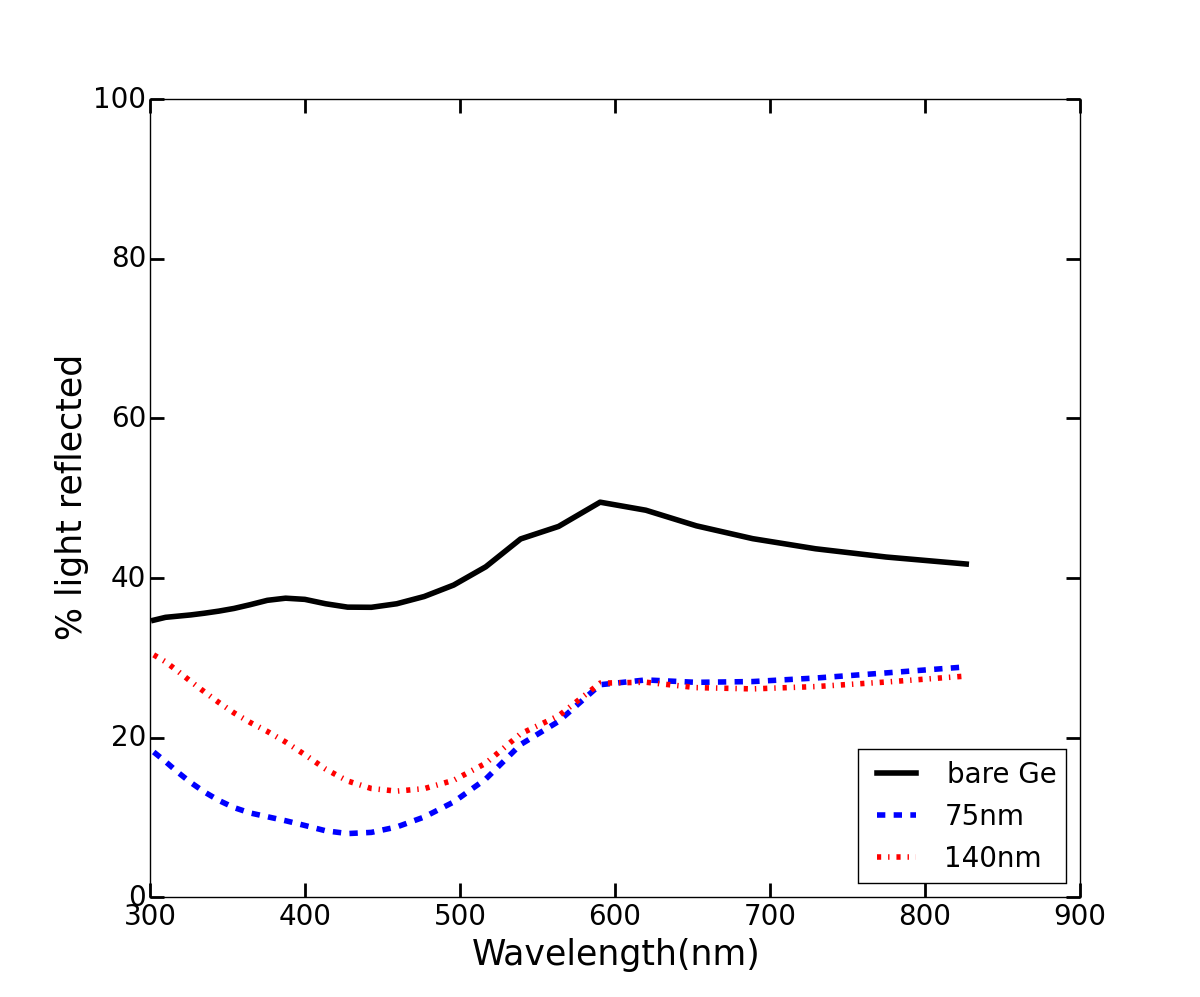}}
\subfigure[$\theta_\text{inc}=15^\circ$]{\includegraphics[scale=0.23]{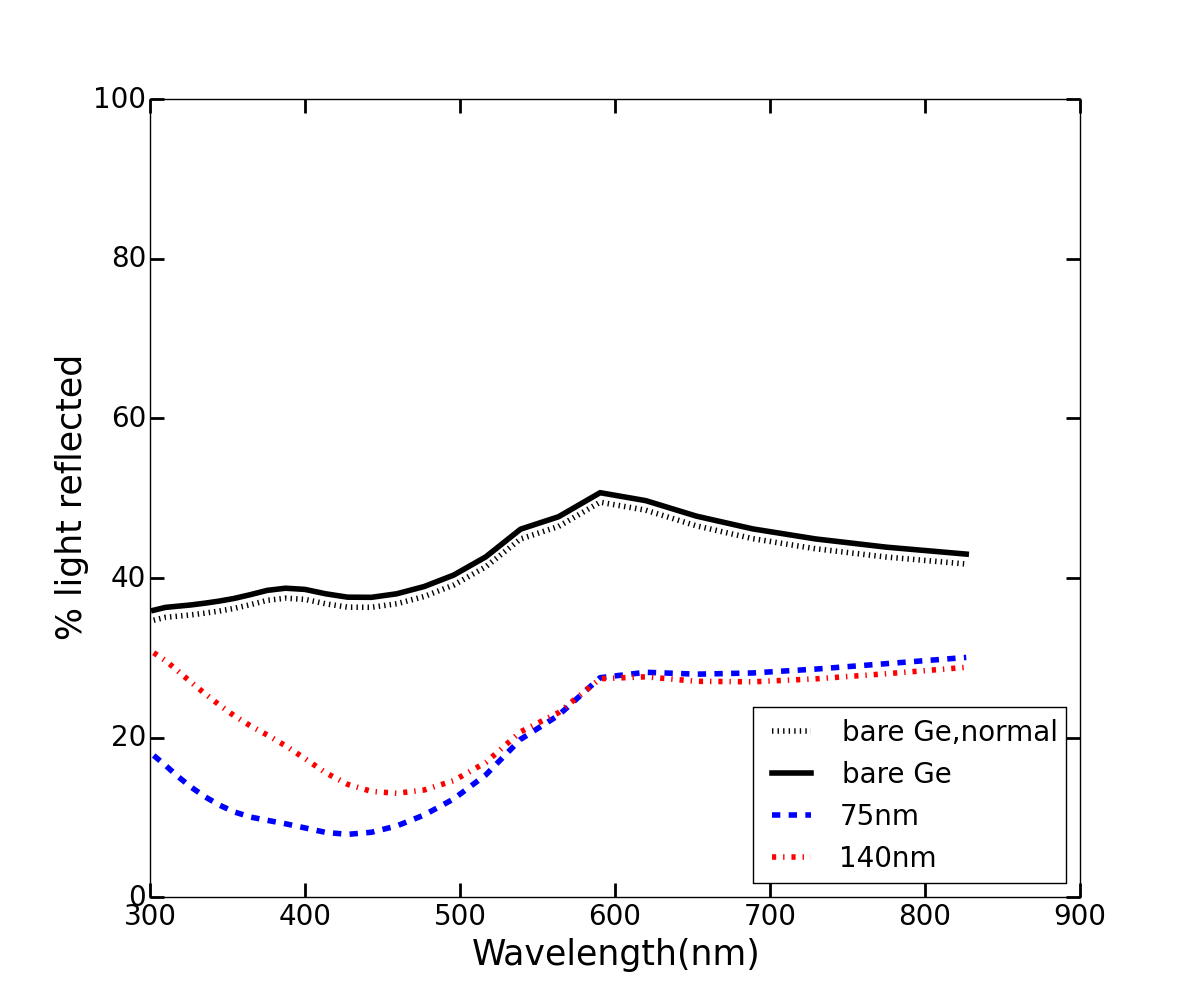}}
\subfigure[$\theta_\text{inc}=45^\circ$]{\includegraphics[scale=0.23]{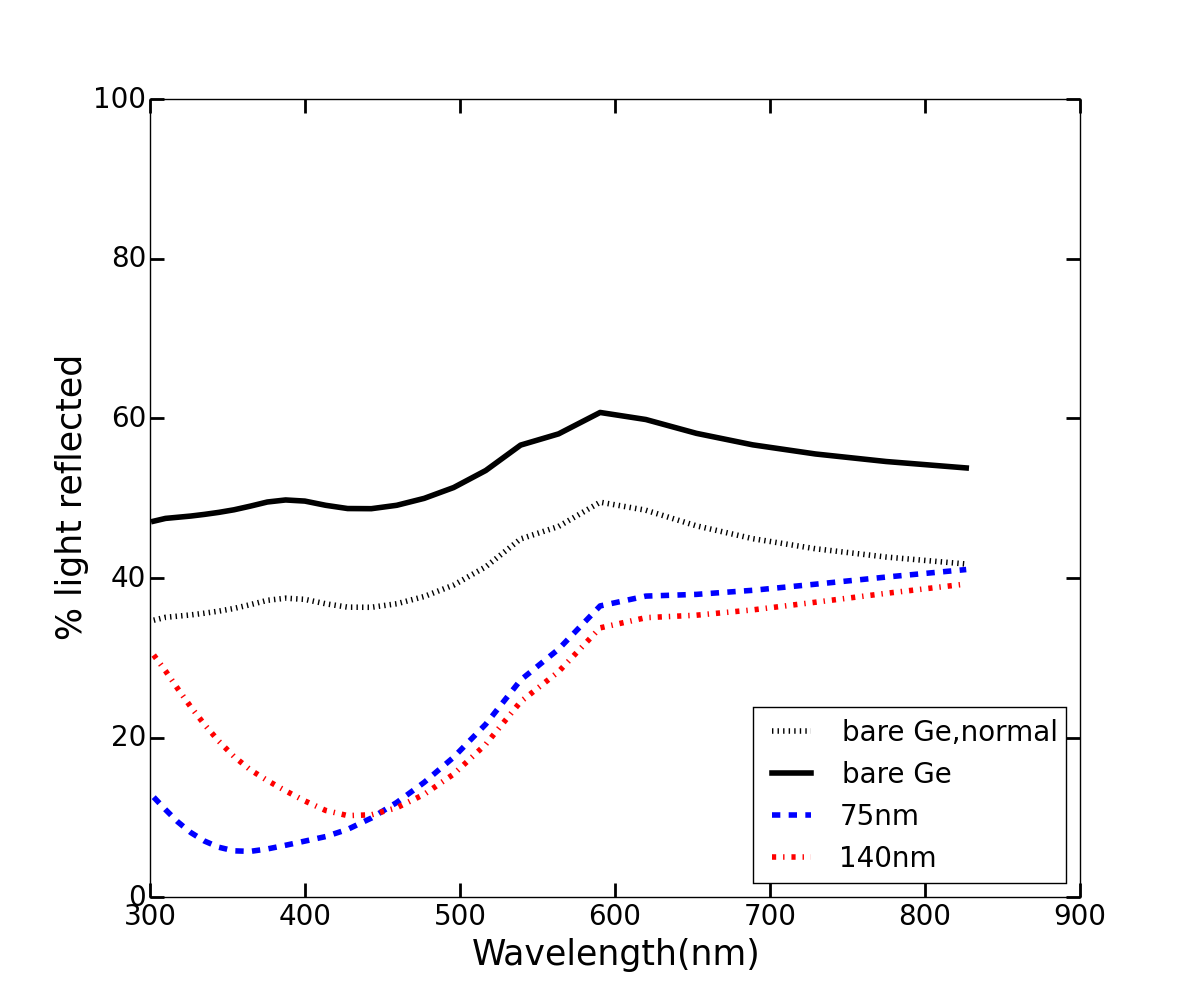}\label{example45}}
\subfigure[$\theta_\text{inc}=85^\circ$]{\includegraphics[scale=0.23]{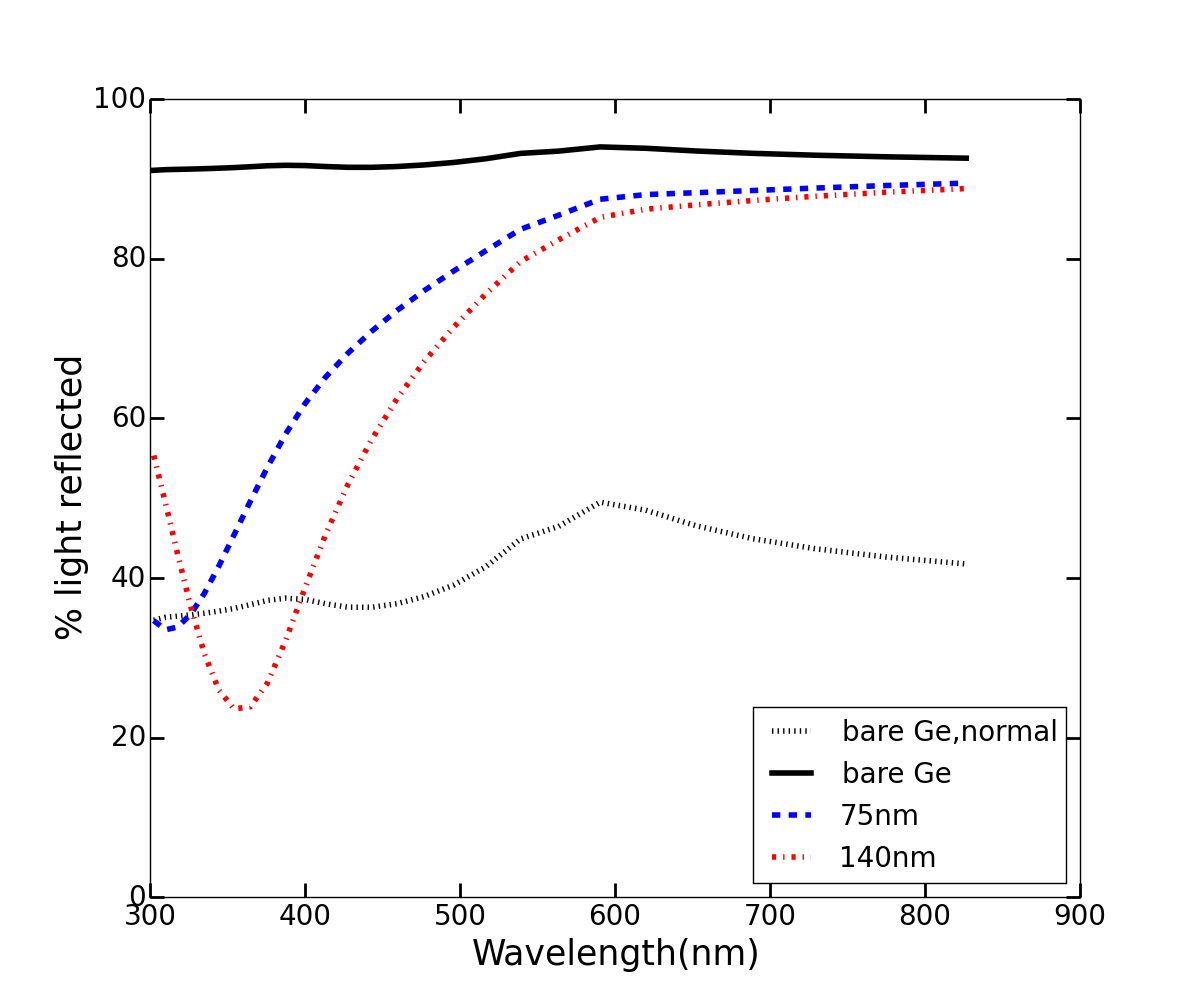}}
\caption[]{Predicted reflectivity values for SiO$_2$ coating on Ge substrate for varied angles of incidence. Black solid curve indicates reflectivity of bare Si~\cite{Aspnes1983DielectricEV}. Angles of incidence are as follows: (a) 0$^\circ$, (b) 15$^\circ$, (c) 45$^\circ$, (d) 85$^\circ$.  Colored reflectivity curves are calculated based on ellipsometric measurements for $n(\lambda)$ (see Figure \ref{NCurves}). Thicknesses of 75 and \unit[140]{nm} (with MSE values 61.05 and 65.02 respectively) are displayed. For plots (b)-(d), the reflectivity curve for normal incidence is provided as a reference (black dotted). For all angles of incidence and thicknesses, coating with SiO$_2$ should improve reflectivity. Recommendations for wavelengths of interest can be found in Section \ref{PredRef}.
\label{RefCurvesSiO2onGe}}
\end{center}
\end{figure}

\section{Predicted Reflectance} \label{PredRef}
From the indexes characterized in Figure \ref{NCurves}, it is possible to reproduce the reflectivity curves for each system. The indices are used as input to equations \ref{ref_first}-\ref{ref_last} and plotted for various thicknesses and angles of incidence\footnote{Incident angles are plotted up to the ellipsometer maximum of 85$^\circ$, although calculations suggest the true maximum could be larger.}; the results for Si are shown in Figure \ref{RefCurvesSiO2onSi}; the results for Ge are shown in Figure~\ref{RefCurvesSiO2onGe}. All curves are plotted in reference to the reflectance for a bare substrate at normal incidence, and the calculation assumes all reflections are specular and there are no losses in the SLAR ($k_{\text{SLAR}}=0$).

For the peak scintillation wavelengths, \unit[645]{nm} and \unit[610]{nm} for ZeSe and ZnMoO$_4$, it is clear that Si substrate reflectivity can be improved with any of the tested SLAR thicknesses: at normal incidence, a \unit[90]{nm} coating of SiO$_2$ improves reflectivity of an Si substrate by $\sim$25\%. The performance of the system depends heavily on the film thickness: \unit[180]{nm} films do not show nearly the same decrease in reflectivity except at higher angles; see Figure~\ref{RefCurvesSiO2onSi}. In contrast, the Ge substrate performance does not depend as heavily on the SLAR thickness. Improvements of up to 22\% were seen for normal incidence for both \unit[75]{nm} and \unit[140]{nm} SiO$_2$ coatings on Ge. For the scintillation wavelength of ZnSe (\unit[645]{nm}), the predicted improvement is $\sim$20\%, which agrees with the findings of previous work~\cite{Beeman2013CurrentExperiment}.

%E: removed citation for teo2 cutoff wl since it's in the intro
For a signal from Cherenkov light, the solution is quite different. TeO$_2$ has a cutoff wavelength at \unit[350]{nm}. At this wavelength, it is difficult for a  Si substrate to get below 20\% reflectivity, so a Ge substrate is a better choice. These results indicate a Ge substrate with a \unit[70]{nm} SiO$_2$ coating is better than larger thicknesses.

%NEED TO EDIT THIS PART ===================================
As seen in Figures \ref{RefCurvesSiO2onSi} and \ref{RefCurvesSiO2onGe}, results also demonstrate that the performance of these light-collecting, target bolometers is greatly affected by the angle of incident light. Significant amounts of light will be lost simply due to reflectivity: scintillation light incident on Ge from a ZnMoO$_4$ primary bolometer at 45$^\circ$ (in the plane of incidence) will be reflected at a 65\% loss (see Figure \ref{example45}). An antireflective coating of 140nm SiO$_2$ can decrease this loss to only about 35\%. 

%Further work is underway to calculate the angular distribution of light leaving the primary bolometer; this analysis will lend itself to calculating the light gain by the eventual target-bolometer system, a great improvement on the assumptions made by \cite{Beeman2013CurrentExperiment}.

\section{Conclusion}
% In order to establish a reliable methodology, allowing a more rigorous recommendation of antireflective coatings, in terms of its composition and thickness, it is necessary to:
% 1) have an integrated way of treating the results taking into account the light emission spectrum of the scintillator and the distribution of the light incidence angles at the surface of the 2nd detector, in order to combine the 4 discrete measurements.
% 2) make final measurements with the detector modules in order to validate and check the accuracy of the methodology that is proposed.
Anti-reflective coatings can significantly increase the efficiency of light collection bolometers in rare event searches with the purpose of reducing background rates. Several coatings were manufactured successfully on  Ge and Si substrates and characterized at room temperature using variable angle ellipsometry. Full characterization of the more rare HfO$_2$ and TiO$_2$ coatings was unsuccessful. Preliminary calculations for both Si and Ge substrates confirm a decrease in reflectivity from SiO$_2$ coatings of various thicknesses at various angles of incidence, an improvement on the calculations performed by \cite{Beeman2013CurrentExperiment} which assumed normal incidence. For the ZnSe and ZnMoO$_4$ scintillation wavelengths of \unit[645]{nm} and \unit[610]{nm}, coatings of \unit[90]{nm} SiO$_2$  on Si or \unit[140]{nm} SiO$_2$ on Ge are recommended. For Cherenkov light from TeO$_2$ at \unit[350]{nm}, a Ge substrate with a \unit[70]{nm} SiO$_2$ coating is recommended. Work is ongoing to fully characterize all coatings at operating and room temperatures, as well as to explore the potential of novel coating techniques.  Additional analysis of the impact of the coatings on system radiopurity and testing of the coatings characteristic bolometric operating temperatures will fully establish the value of SLAR coatings in rare event searches. 

\subsection*{Acknowledgments}
This work was funded by the Hellman Fellow program at UCLA. Special thanks to the UCLA NanoLab (Max Ho, Tom Lee, Wilson Lin) and UCSB NanoLab (Tom Reynolds, Brian Thibeault). Many thanks also to Matteo Biassoni, Huan Huang, Jacob Feintzeig, Laura Gladstone, Jon Ouellet, and other CUORE collaborators for their valuable insight and discussion. A special thank you to Laura Gladstone and Jacob Siegel for editing the manuscript.

\bibliographystyle{JHEP}
\bibliography{ARC_Paper.bib}

\end{document}